\documentclass[12pt]{article}
																																																																																											\pdfoutput=1																																																																																			 
\usepackage{amsmath,amssymb}
\usepackage{graphicx}
\usepackage[pdftex]{hyperref}
\DeclareGraphicsExtensions{.eps,.bmp,.wmf,.jpeg,.pdf}
\numberwithin{equation}{section}

\def\be{\begin{equation}}
\def\ee{\end{equation}}

\def\bea{\begin{eqnarray}}
\def\eea{\end{eqnarray}}
\title{Big Rip and Little Rip solutions in scalar model with kinetic and Gauss Bonnet couplings}
\author{L. N. Granda\thanks{ngranda@univalle.edu.co}\, and \ E. Loaiza\thanks{edwin.loaiza@correounivalle.edu.co, Sede Buga}\\ {\small\it Departamento de Fisica, Universidad del Valle}\\{\small\it A.A. 25360, Cali, Colombia}} 
\date{}
\begin{document}
\maketitle

\begin{abstract}
\noindent Late time cosmological solutions for scalar field model with kinetic and Gauss Bonnet couplings are considered. The quintom scenario is realized with and without Big Rip singularity. We find that under specific choice of the Gauss Bonnet coupling, the model considerable simplifies, giving rise to solutions where the kinetic term is proportional to the square of the Hubble parameter. This allows to reconstruct the model for a suitable cosmological evolution. We considered a solution that matches the observed behavior of the equation of state, while Big Rip singularity may be present or absent, depending on the parameters of the solution. Evolutionary scenarios known as Little Rip, have also been considered.

\noindent PACS 98.80.-k, 95.36+x, 04.50.kd
\end{abstract}

\section{Introduction}
\noindent 
Recent observational data provide strong evidence for an accelerated expanding universe at the current epoch \cite{riess}, \cite{perlmutter}, \cite{hicken}, \cite{komatsu}, \cite{percival}. These astrophysical observations also suggest that the universe is spatially flat with high accuracy, and is composed of about 27\% of dark matter (including the usual baryonic matter),  and 73\% of homogeneously distributed new type of negative pressure matter, called dark energy, which causes the current accelerated expansion.
Among the variety of dynamical approaches to the dark energy (DE) problem (see \cite{copeland}, \cite{sahnii}, \cite{padmanabhan}), the scalar tensor theories which interpolate between the scalar field and modified gravity models, allow in principle to evade the coincidence problem, and in some cases give rise to quintom behavior \cite{peri}, \cite{maeda}.
In the present work, we consider a string and higher-dimensional
gravity inspired scalar field model, with two kind of couplings: kinetic coupling to curvatures and Gauss Bonnet coupling to the scalar field. The GB term is topologically invariant in four dimensions and does not contribute to the equations of motion. Nevertheless it affects cosmological dynamics when it is coupled to a dynamically evolving scalar field through arbitrary function of the field, giving rise to second order differential equations of motion, preserving the theory ghost free. Therefore, the coupled GB term seems as a natural generalization of the scalar field with non-minimal kinetic coupling to curvature \cite{granda,granda1}.\\
\noindent Some late time cosmological aspects of scalar field model with derivative couplings to curvature have been considered in \cite{sushkov}, \cite{granda,granda1,granda2}, \cite{gao}. On the other hand, the GB invariant coupled to scalar field have been extensively studied. In \cite{sergei12} the GB correction was proposed to study the dynamics of dark energy, where it was found that quintessence or phantom phase may occur in the late time universe. Accelerating cosmologies with GB correction in four and higher dimensions have been discussed in \cite{tsujikawa}, \cite{leith}, \cite{maartens}. The modified GB theory applied to dark energy have been suggested in \cite{sergei14}, and different aspects of the modified GB model applied to late time acceleration, have been considered among others, in \cite{sergeio1}, \cite{sergeio2}, \cite{sergei15}, \cite{carter}, \cite{tretyakov}.\\
All these studies demonstrate that it is quite plausible that the scalar-tensor couplings predicted by the fundamental theory may become important at current, low-curvature
universe (see \cite{sergeio3} and \cite{sergei11} for review).\\
The most general second-order ghost-free scalar-tensor Lagrangian with couplings to curvature, can be originated from toroidal compactification of $4+N$ dimensional Lagrangian of pure gravity, as shown in \cite{amendola}. In this compactification the scalar field plays the role of the overall size of the $N$-torus, and the couplings appear as exponentials of the scalar field. This general Lagrangian also appears in the next to leading order corrections in the $\alpha'$ expansion of the string theory \cite{tseytlin}, \cite{cartier}. 
In the present study we consider those terms in this ghost-free scalar-tensor Lagrangian that are coupled to the curvature, where we use more general couplings in order to reconstruct appropriate cosmological scenarios satisfying current astrophysical observations. 
A solution unifying early time decelerated behavior, with late time accelerated and phantom phases will be studied. We considered solutions with and without
Big Rip singularity. The equation of state presents minimum at the future, bellow the phantom divide, turning back asymptotically to de Sitter phase. New solutions
known as Little Rip, have also been considered.
In section II we introduce the model and give the general equations, which are then expanded on the FRW metric. In section III and IV we present solutions for the Hubble parameter with Big Rip and Little Rip singularities, and reconstruct the model according to these cosmological scenarios. Concluding remarks are given in section V.

\section{Field Equations}
Let us start with the  action for scalar field with kinetic terms non-minimally coupled to curvature and coupled to Gauss Bonnet (GB) curvature
\be\label{eq1}
\begin{aligned}
S=&\int d^{4}x\sqrt{-g}\Big[\frac{1}{16\pi G} R -\frac{1}{2}\partial_{\mu}\phi\partial^{\mu}\phi+F_1(\phi)G_{\mu\nu}\partial^{\mu}\phi\partial^{\nu}\phi- V(\phi)+F_2(\phi){\cal G}\Big]
\end{aligned}
\ee
\noindent where $G_{\mu\nu}=R_{\mu\nu}-\frac{1}{2}g_{\mu\nu}R$, ${\cal G}$ is the 4-dimensional GB invariant ${\cal G}=R^2-4R_{\mu\nu}R^{\mu\nu}+R_{\mu\nu\rho\sigma}R^{\mu\nu\rho\sigma}$. The coupling $F_1(\phi)$ has dimension of $(length)^2$, and the coupling $F_2(\phi)$ is dimensionless. Besides the couplings of curvatures with kinetic terms, one may expect that the presence of GB coupling term may be relevant for the explanation of dark energy phenomena. The GB coupling has the advantage that does not make contributions higher than second order (in the metric) to the equations of motion, and therefore does not introduce ghost terms into the theory. Hence, the equations derived from this action contain only second derivatives of the metric and the scalar field, avoiding problems with higher order derivatives \cite{capozziello1}.\\
Taking the variation of action (\ref{eq1}) with respect to the metric, we obtain a general expression of the form 
\be\label{eq2}
R_{\mu\nu}-\frac{1}{2}g_{\mu\nu}R=\kappa^2 T_{\mu\nu}
\ee
where $\kappa^2=8\pi G$, $T_{\mu\nu}^m$ is the usual energy-momentum tensor for matter component, the tensor $T_{\mu\nu}$ represents the variation of the terms which depend on the scalar field $\phi$ and can be written as
\be\label{eq3}
T_{\mu\nu}=T_{\mu\nu}^{\phi}+T_{\mu\nu}^{K}+T_{\mu\nu}^{GB}
\ee
where $T_{\mu\nu}^{\phi}$,  correspond to the variations of the standard minimally coupled terms, $T_{\mu\nu}^{K}$ comes from the kinetic coupling, and $T_{\mu\nu}^{GB}$ comes from the variation of the coupling with GB. 
Due to the kinetic coupling with curvature and the GB coupling, the quantities derived from this energy-momentum tensors will be considered as effective ones. The variations are given by \cite{granda20,granda21}
\be\label{eq4}
T_{\mu\nu}^{\phi}=\nabla_{\mu}\phi\nabla_{\nu}\phi-\frac{1}{2}g_{\mu\nu}\nabla_{\lambda}\phi\nabla^{\lambda}\phi
-g_{\mu\nu}V(\phi)
\ee
\be\label{eq5}
\begin{aligned}
T_{\mu\nu}^{K}=&\left(R_{\mu\nu}-\frac{1}{2}g_{\mu\nu}R\right)F_1(\phi)\nabla_{\lambda}\phi\nabla^{\lambda}\phi+g_{\mu\nu}\nabla_{\lambda}\nabla^{\lambda}\left(F_1(\phi)\nabla_{\gamma}\phi\nabla^{\gamma}\phi\right)\\
&-\frac{1}{2}(\nabla_{\mu}\nabla_{\nu}+\nabla_{\nu}\nabla_{\mu})\left(F_1(\phi)\nabla_{\lambda}\phi\nabla^{\lambda}\phi\right)+R F_1(\phi)\nabla_{\mu}\phi\nabla_{\nu}\phi\\& -2F_1(\phi)\left(R_{\mu\lambda}\nabla^{\lambda}\phi\nabla_{\nu}\phi+R_{\nu\lambda}\nabla^{\lambda}\phi\nabla_{\mu}\phi\right)+g_{\mu\nu}R_{\lambda\gamma}F_1(\phi)\nabla^{\lambda}\phi\nabla^{\gamma}\phi\\
&+\nabla_{\lambda}\nabla_{\mu}\left(F_1(\phi)\nabla^{\lambda}\phi\nabla_{\nu}\phi\right)+\nabla_{\lambda}\nabla_{\nu}\left(F_1(\phi)\nabla^{\lambda}\phi\nabla_{\mu}\phi\right)\\
&-\nabla_{\lambda}\nabla^{\lambda}\left(F_1(\phi)\nabla_{\mu}\phi\nabla_{\nu}\phi\right)-g_{\mu\nu}\nabla_{\lambda}\nabla_{\gamma}\left(F_1(\phi)\nabla^{\lambda}\phi\nabla^{\gamma}\phi\right)
\end{aligned}
\ee
and 
\be\label{eq7a}
\begin{aligned}
T_{\mu\nu}^{GB}=&4\Big([\nabla_{\mu}\nabla_{\nu}F_2(\phi)]R-g_{\mu\nu}[\nabla_{\rho}\nabla^{\rho}F_2(\phi)]R-2[\nabla^{\rho}\nabla_{\mu}F_2(\phi)]R_{\nu\rho}-2[\nabla^{\rho}\nabla_{\nu}F_2(\phi)]R_{\nu\rho}\\
&+2[\nabla_{\rho}\nabla^{\rho}F_2(\phi)]R_{\mu\nu}+2g_{\mu\nu}[\nabla^{\rho}\nabla^{\sigma}F_2(\phi)]R_{\rho\sigma}-2[\nabla^{\rho}\nabla^{\sigma}F_2(\phi)]R_{\mu\rho\nu\sigma}\Big)
\end{aligned}
\ee
In this last expression the properties of the 4-dimensional GB invariant have been used (see \cite{sergei12}, \cite{farhoudi}).
Variating with respect to the scalar field gives the equation of motion
\be\label{eq7}
\begin{aligned}
&-\frac{1}{\sqrt{-g}}\partial_{\mu}\left[\sqrt{-g}\left(R F_1(\phi)\partial^{\mu}\phi-2R^{\mu\nu}F_1(\phi)\partial_{\nu}\phi+\partial^{\mu}\phi\right)\right]+\frac{dV}{d\phi}+\\
&\frac{dF_1}{d\phi}\left(R\partial_{\mu}\phi\partial^{\mu}\phi-2 R_{\mu\nu}\partial^{\mu}\phi\partial^{\nu}\phi\right)-\frac{dF_2}{d\phi}{\cal G}=0
\end{aligned}
\ee
Considering the spatially-flat Friedmann-Robertson-Walker (FRW) metric,
\be\label{eq8}
ds^2=-dt^2+a(t)^2\left(dr^2+r^2d\Omega^2\right)
\ee
And assuming an homogeneous time-depending scalar field $\phi$ , the $(00)$ and $(11)$ components of the Eq. (\ref{eq2}), from (\ref{eq3}-\ref{eq7a}) take the form (with the Hubble parameter $H=\dot{a}/a$)
\be\label{eq9}
H^2=\frac{\kappa^2}{3}\rho_{eff}=\frac{\kappa^2}{3}\left(\frac{1}{2}\dot{\phi}^2+V(\phi)+9H^2F_1(\phi)\dot{\phi}^2-24H^3\frac{dF_2}{d\phi}\dot{\phi}\right)
\ee
and
\be\label{eq13}
\begin{aligned}
&-2\dot{H}-3H^2=\kappa^2 p_{eff}=\kappa^2\Big[\frac{1}{2}\dot{\phi}^2-V(\phi)-\left(3H^2+2\dot{H}\right)F_1(\phi)\dot{\phi}^2\\&-2 H\left(2F_1(\phi)\dot{\phi}\ddot{\phi}+\frac{dF_1}{d\phi}\dot{\phi}^3\right)
+8H^2\frac{dF_2}{d\phi}\ddot{\phi}+8H^2\frac{d^2F_2}{d\phi^2}\dot{\phi}^2+16H\dot{H}\frac{dF_2}{d\phi}\dot{\phi}+16H^3\frac{dF_2}{d\phi}\dot{\phi}\Big]
\end{aligned}
\ee
I the present study we have assumed scalar field dominance (the matter term is absent in the action). The equation of motion for the scalar field (\ref{eq7}) takes the form
\be\label{eq14}
\begin{aligned}
&\ddot{\phi}+3H\dot{\phi}+\frac{dV}{d\phi}+3 H^2\left(2F_1(\phi)\ddot{\phi}+\frac{dF_1}{d\phi}\dot{\phi}^2\right)
+18 H^3F_1(\phi)\dot{\phi}+\\
&12 H\dot{H}F_1(\phi)\dot{\phi}-24\left(\dot{H}H^2+H^4\right)\frac{dF_2}{d\phi}=0
\end{aligned}
\ee
where the first three terms correspond to the minimally coupled field.\\
\noindent Assuming an asymptotic behavior of the scalar field as $\phi=\phi_0=const.$, and $F_2(\phi)=const.$, then independently of $F_1(\phi)$ he model presents de Sitter solution, as can be seen from Eqs. (\ref{eq9}) and (\ref{eq14}). From (\ref{eq9}) and (\ref{eq14}) it follows that $V=V_0=const$ and $H=H_0=\kappa\sqrt{V_0/3}$.\\
In the following sections we study cosmological solutions of Eqs. (\ref{eq9}) and (\ref{eq14}) giving rise to accelerated expansion,
including quintom behavior, and presenting Big Rip and Little Rip singularities.
\section{Solution with and without Big Rip singularity}
We continue studying some solutions to Eqs. (\ref{eq9}) and (\ref{eq14}), in the important case when the scalar field potential is absent (i.e. $V=0$), which leaves the two couplings $F_1(\phi)$ and $F_2(\phi)$ as the degrees of freedom that describe the cosmological dynamics. Making $\dot{\phi}^2=\psi$, from Eq. 
(\ref{eq9}) with $V=0$, it follows
\be\label{eq14a}
F_1\psi=\frac{1}{3}-\frac{\psi}{18 H^2}+\frac{8}{3}H\frac{dF_2}{dt}
\ee
replacing in (\ref{eq14}) after simplifications, we obtain
\be\label{eq14b}
H\frac{d\psi}{dt} +6H^2\psi-\psi\frac{dH}{dt}+48H^3\frac{dH}{dt}\frac{dF_2}{dt}+72H^5\frac{dF_2}{dt}+24H^4\frac{d^2F_2}{dt^2}+12H^2\frac{dH}{dt}+18H^4=0
\ee
here we have set $\kappa^2=1$. Let's consider the GB coupling as proposed in \cite{granda21}
\be\label{eq14c}
\frac{dF_2(t)}{dt}=\frac{g}{H(t)}
\ee
Replacing (\ref{eq14c}) in (\ref{eq14b}) one obtains
\be\label{eq14d}
H\frac{d\psi}{dt}+6H^2\psi-\psi\frac{dH}{dt}+12(2g+1)H^2\frac{dH}{dt}+18(4g+1)H^4=0
\ee
An interesting and viable late time cosmological solution to this equation is the following
\be\label{eq15}
H(t)=\frac{p}{t}+\frac{\alpha t+\beta}{\gamma t+\eta}
\ee
where $p,\beta,\eta$ are dimensionless constants, and $\alpha, \gamma$ are constants with inverse time dimension.
At early times the first term dominates, giving rise to power-law behavior $H\sim p/t$. At late times the second term offers an interesting alternative to the cosmological constant, as it behaves as $\alpha/\gamma$, which is approximately the current cosmological behavior. Depending on the sign of $\gamma$ and $\eta$ may present Big Rip singularity. Integrating with respect to time, one finds the scale parameter as
\be\label{eq15a}
a(t)=t^p e^{\alpha t/\gamma}(\gamma t+\eta)^{\frac{\beta}{\gamma}-\frac{\alpha\eta}{\gamma^2}}
\ee
Note that $a=0$ at $t=0$, indicating that the solution can cover the inflationary epoch.
Replacing in (\ref{eq14d}) one obtains
\be\label{eq16}
\psi(t)=6\left(\frac{p}{t}+\frac{\alpha t+\beta}{\gamma t+\eta}\right)^2
\ee
where we have simplified the solution by setting the integration constant equal to zero. Note that in this specific case $\psi(t)=6H(t)^2$. Integrating the square root of $\psi$ we find the scalar field
\be\label{eq17}
\phi(t)=\sqrt{6}\left(\frac{\alpha t}{\gamma}+p\ln(t)+\frac{(\beta\gamma-\alpha\eta)\ln(\gamma t+\eta)}{\gamma^2}\right)
\ee
Note that if the parameters in (\ref{eq15}) are related by $\beta\gamma-\alpha\eta=0$, then the scalar field simplifies and the Hubble function takes the form $H=p/t+\beta/\eta$, which has been already considered \cite{sergei17}.
The GB coupling is obtained from (\ref{eq14c})
\be\label{eq18}
\begin{aligned}
F_2(t)=&-\frac{3}{4}\left[\frac{\gamma t}{\alpha}+\frac{(\gamma A-\alpha\beta\eta-3p\alpha\gamma\eta)}{\alpha^2\sqrt{4p\alpha\eta-A}}\arctan\left[\frac{2\alpha t+\gamma p +\beta}{\sqrt{4p\alpha\eta-A}}\right]\right]\\
&-\frac{3}{8\alpha^2}(\alpha\eta-p\gamma^2-\beta\gamma)\ln\left(\alpha t^2+(\beta+p\gamma)t+p\eta\right)
\end{aligned}
\ee
where $A=\beta^2+2p\beta\gamma+p^2\gamma^2$. And the kinetic coupling from (\ref{eq14a}) is  
\be\label{eq18a}
F_1(t)=-\frac{1}{3}\left(\frac{p}{t}+\frac{\alpha t+\beta}{\gamma t+\eta}\right)^{-2}
\ee
The effective EoS is obtained from the solution (\ref{eq15})
\be\label{eq19}
w_{eff}=-1+\frac{2[(p\gamma^2+\beta\gamma-\alpha\eta)t^2+2p\gamma\eta t+p\eta^2]}{3(\alpha t^2+(p\gamma+\beta)t+p\eta)^2}
\ee
This EoS has two important limits:
\be\label{eq20}
\lim_{t\to 0}w_{eff}(z)=-1+\frac{2}{3p}
\ee
the particular value $p=2/3$ gives the known matter dominance regime. At late times, according to Eq. (\ref{eq15}) the universe becomes dominated by the cosmological constant and ends ($t\longrightarrow\infty$) in a de Sitter phase $w_{eff}=-1$. We can use the parameters appearing in (\ref{eq15}) to meet the behavior demanded by the
current astrophysical observations. Normalizing the cosmological time so that the current time corresponds to $t=t_0=1$, and taking the current value of the EoS $w_{eff}(1)=w_0=-1$ (which indeed, according to observations could be a very close to $-1$) we find the following restriction on the parameters
\be\label{eq21}
p(\gamma+\eta)^2+\beta\gamma-\alpha\eta=0
\ee
we can also normalize the scale parameter $a(t)$ so that $a(1)=1$. This gives the restriction
\be\label{eq22}
\beta=\alpha\left(\frac{\eta}{\gamma}-\frac{1}{\ln(\eta+\gamma)}\right)
\ee
these two restrictions allow to express the EoS in terms of the three parameters $p$, $\eta$ and $\gamma$ as follows
\be\label{eq22a}
w_{eff}=-1-\frac{2\gamma^4\eta((2\gamma+\eta)t+\eta)(t-1)}{3p\left[(\gamma+\eta)^2(\gamma t+\eta)t\log(\gamma+\eta)+\gamma\eta(\gamma-(2\gamma+\eta)t)\right]^2}
\ee
The transition from decelerated to accelerated phase takes place when $w_{eff}(t_{tr})=-1/3$, which gives the following expression to find $t_{tr}$ for given $p$, $\gamma$ and $\eta$
\be\label{eq22b}
p\left[\eta\gamma^2-\eta\gamma(2\gamma+\eta)t_{tr}+(\gamma+\eta)^2\ln(\gamma+\eta)(\gamma t_{tr}+\eta)t_{tr}\right]^2=\gamma^4\eta\left((2\gamma+\eta)t_{tr}+\eta\right)(1-t_{tr})
\ee
where we replaced $\alpha$ and $\beta$ from (\ref{eq21}) and (\ref{eq22}). There is one more phase presented in this solution, which according to the restriction (\ref{eq21}) is taking place currently: the transition to phantom phase delimited by $w_0=-1$. Then we can expect that at $t>1$ the EoS takes values $w_{eff}<-1$. An important feature of the EoS is that it does not decrease monotonically with time, but has a minimum at $t_{min}$ satisfying 
\be\label{eq23}
\ln(\gamma+\eta)\left(\gamma(2\gamma+\eta)t_{min}^3-3\gamma^2t_{min}^2-3\gamma\eta t_{min}-\eta^2\right)+\gamma\eta=0
\ee
As we will see in a numerical example, this minimum occurs bellow the phantom barrier, but then the EoS starts increasing towards the value $w_0=-1$ at $t\rightarrow\infty$.
In the solution (\ref{eq15}) we can distinguish two important cases:  if we consider that $\gamma$ and $\eta$ are positive, then the cosmological evolution avoid Big Rip singularities, and on the contrary, if we consider $\gamma<0$, then the late time evolution faces Big Rip singularity.
Taking for instance $p=2/3$, and using (\ref{eq21}) and (\ref{eq22}), we define the behavior of the EoS so that at $t=1$ crosses the phantom barrier, where $\gamma$ and $\eta$ are used to properly determine the time of transition deceleration-acceleration, according to Eq. (\ref{eq22a}). Taking for instance the values $\gamma=0.93$ and $\eta=0.8$, then using Eqs. (\ref{eq21})-(\ref{eq23}) we find $t_{tr}\sim 0.64$ and $t_{min}\sim 1.5$. 
In Fig. 1 we show the behavior of the EoS in this case
\begin{center}
\includegraphics [scale=0.7]{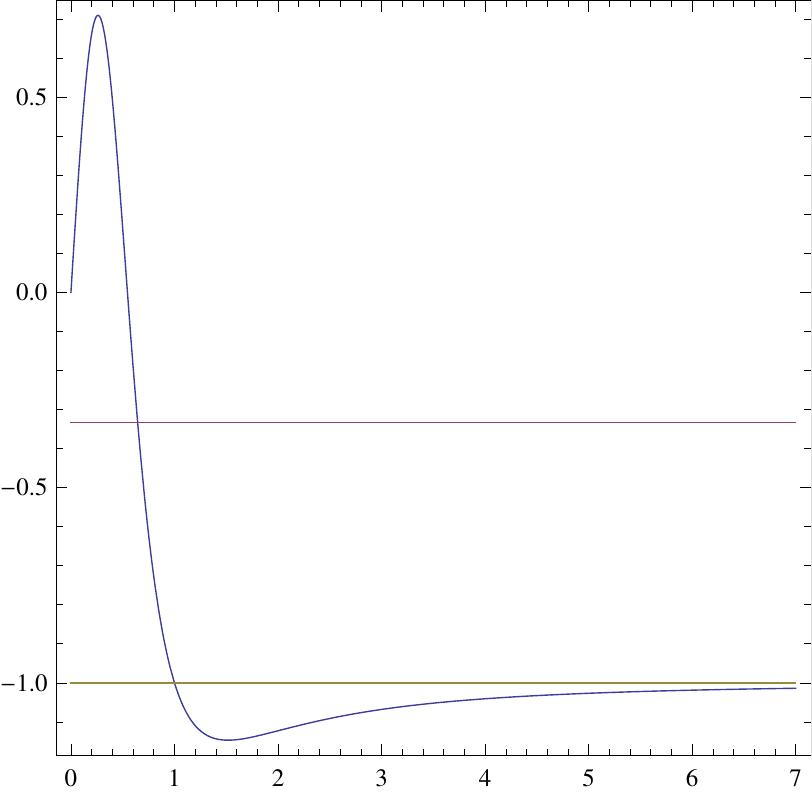}
\end{center} 
\begin{center}
{Fig. 1 \it The effective EoS for $p=2/3,\gamma=0.93,\eta=0.8$. Note the presence of the three phases: matter dominance (decelerated phase) above the line $w_{eff}=-1/3$, the accelerated-quintessence phase between the lines $w_{eff}=-1/3$ and $w_{eff}=-1$, and phantom phase bellow the line $w_{eff}=-1$}.
\end{center}
An important aspect of this solution is that despite the fact that the universe enters in the phantom phase, nevertheless the effective energy density remains finite, as can be seen by the behavior of the Hubble function (\ref{eq15}). Similar behavior for the EoS was obtained in a quintom model with spinor field \cite{yifu}.\\
\noindent{\bf Big Rip Singularity}\\
According to solution (\ref{eq15}) the future Big Rip singularity occurs provided that $\gamma<0$. Thus, we define $t_s=-\eta/\gamma$ as the time when the Big rip singularity is approached. From (\ref{eq15}) it follows that as $t\rightarrow t_s$, $H\rightarrow\infty$ and $\dot{H}\rightarrow\infty$, meaning that $\rho_{eff}\rightarrow\infty$, $p_{eff}\rightarrow\infty$, and from (\ref{eq15a}) $a\rightarrow\infty$. In Fig. 2 we show the EoS for different cases with Big Rip singularity, showing different behaviors of the EoS depending on $\gamma$ and $\eta$
\begin{center}
\includegraphics [scale=0.7]{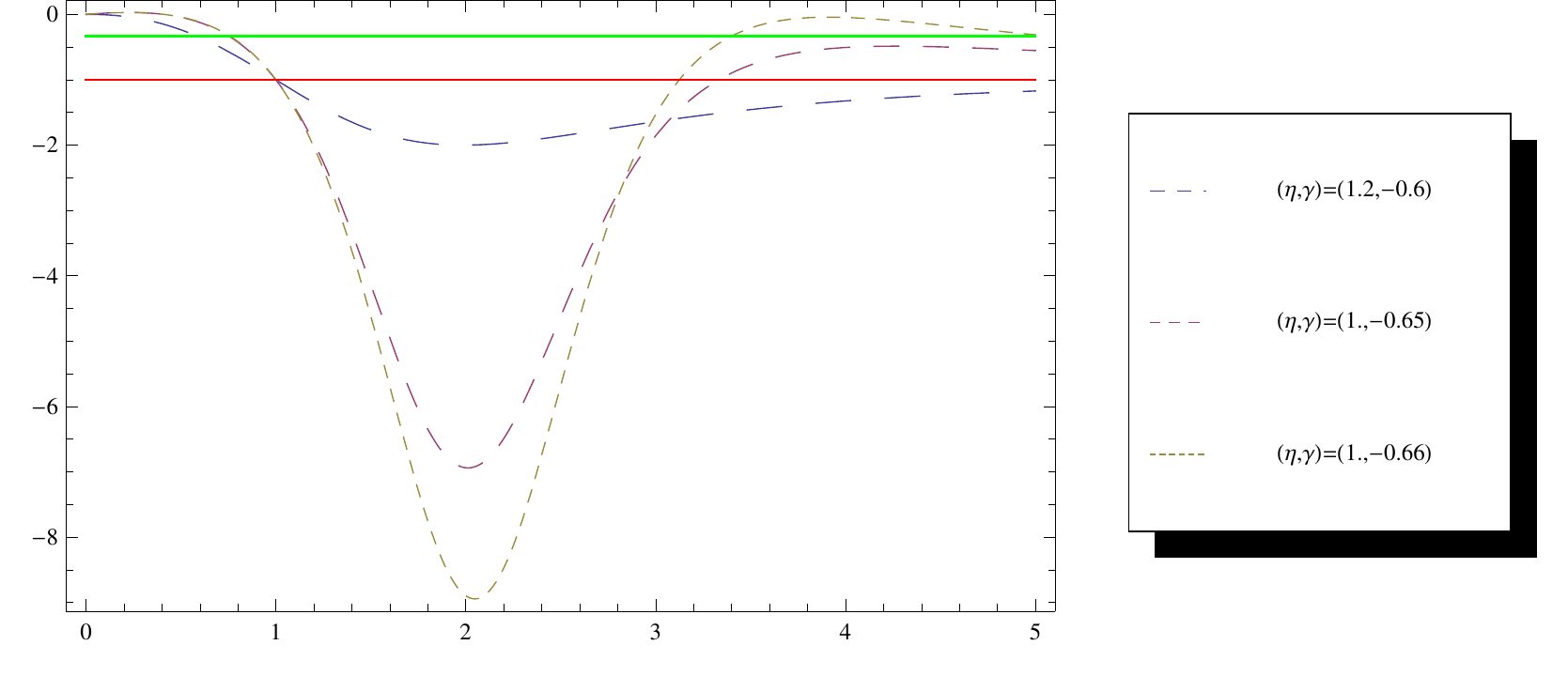}
\end{center} 
\begin{center}
{Fig. 2 \it The effective EoS for $p=2/3$ and different values of $(\eta,\gamma)$:$(1.2,-0.6)$, $(1,-0.65)$, $(1,-0.66)$. Despite the Big Rip singularity, the EoS has a smooth behavior. The curve $(1,-0.66)$ returns back to the decelerated phase, entering again in the accelerated-quintessence phase at far future. Note that the Big Rip singularity may be delayed by making $t_s=-\eta/\gamma$ bigger.}
\end{center}
Therefore, the Hubble parameter given by (\ref{eq15}) gives rise to Big Rip (BR) singularity (Fig. 2), depending on the values of $\gamma$ and $\eta$. Thus if $\eta=-3\gamma$, then the BR singularity occurs at $t_s=3$, i.e. approximately  within 28 Gys. from now.\\
Is worth to note that the only future singularity in $a$, $H$ and $\dot{H}$ occurs simultaneously at $t\rightarrow t_s=-\eta/\gamma$ (with $\gamma<0$), when $\rho\rightarrow\infty$ and $|p|\rightarrow\infty$. This excludes type III singularity according to the classification given in (\cite{sergeio5}), as it requires finite $a_s\neq 0$ at $t\rightarrow t_s$. The type II singularity requires finite $a$ and $\rho$, while $|p|\rightarrow\infty$, which can not be achieved in the present solution. On the other hand, by simultaneously solving the equations $H=0$ and $\dot{H}=0$ (keeping $\gamma<0$) we can find a finite time $t_c$ (in terms of the parameters $\alpha,\beta,\gamma$ and $\eta$) when $\rho\rightarrow 0$ and $|p|\rightarrow 0$, and the EoS diverges. Nevertheless, at this time ($t_c$) the higher order derivatives of $H$ become finite, and therefore there is not singularity of the type IV as described in \cite{sergeio5}.

\section{Little Rip solutions}
In the new solutions known as Little Rip, neither the scale factor nor the energy density become infinity in finite time. As in the BR singularity, such solutions cause the effect of the structure disintegration in finite time, which can be either earlier or later than in a BR model \cite{paul}, \cite{brevik}, \cite{paul1}, \cite{sergeio4}. 
Let's consider the behavior of the Hubble parameter
\be\label{eq23a}
H(t)=H_0 e^{h t}
\ee 
where $h$ is a positive constant. Integrating this equation we find the scale parameter as 
\be\label{eq24}
a(t)=a_0 e^{e^{ht}}
\ee
which leads to the behavior of the density as $\rho\propto (\ln a)^2$, which increases with $a$. From (\ref{eq23a}) it follows that $\dot{H}>0$, which reproduces a super-accelerated phase free of future singularity. The properties of dissolution of structure of this kind of solutions is given in \cite{paul}.\\
In the frame of the present model we may reconstruct the couplings and the scalar field, according to the solution (\ref{eq23}). Replacing this solution in (\ref{eq14d}) one obtains (for the case of $g=-3/4$)
\be\label{eq25}
\psi(t)=6H_0^2 e^{2ht}+C e^{-\frac{6H_0 e^{ht}}{h}+ht}
\ee
where $C$ is the integration constant. The scalar field is given by
\be\label{eq26}
\phi(t)=\int{(6H_0^2 e^{2ht}+C e^{-\frac{6H_0 e^{ht}}{h}+ht})^{1/2}dt} 
\ee
Setting $C=0$, the scalar field takes the simple form $\phi=\sqrt{6}H_0e^{ht}/h$. The GB coupling, as follows from (\ref{eq14c}) is given by
\be\label{eq26a}
F_2(t)=\frac{3}{4H_0h}e^{-h t}
\ee
The kinetic coupling  from (\ref{eq14a}), and in the case $C=0$, becomes
\be\label{eq26b}
F_1(t)=-\frac{1}{3H_0^2}e^{-2ht}
\ee 
In terms of the scalar field the kinetic and GB couplings are given by the expressions
\be\label{eq26c}
F_1(\phi)=-\frac{2}{h^2}\phi^{-2},\,\,\,\,\,\, F_2(\phi)=\frac{3\sqrt{6}}{4h^2}\phi^{-1}
\ee
Recovering the Newtonian coupling ($\kappa^2=8\pi G$), we find that the reconstructed model takes the form
\be\label{eq26d}
\begin{aligned}
S=&\int d^{4}x\sqrt{-g}\Big[\frac{1}{2\kappa^2} R -\frac{1}{2}\partial_{\mu}\phi\partial^{\mu}\phi-\frac{2}{\kappa^2h^2}\frac{1}{\phi^2}G_{\mu\nu}\partial^{\mu}\phi\partial^{\nu}\phi+ \frac{3}{4\kappa^3h^2}\frac{1}{\phi}{\cal G}\Big]
\end{aligned}
\ee
According to \cite{paul, brevik}, the LR solution (\ref{eq23a}) resembles the $\Lambda$CDM at low redshift and is consistent with current supernova
observations (which constrain the parameter $h$).
In fact we can give a dynamical interpretation to a class of solutions that satisfy the requirement of Big Rip or Little Rip (see \cite{paul}-\cite{paul1}), in the frame of the present model with kinetic and GB couplings. First note that the Eq. (\ref{eq14d}) for the fixed $g=-3/4$ takes the form
\be\label{eq27}
H\frac{d\psi}{dt}+\left(6H^2-\frac{dH}{dt}\right)\psi-6H^2\frac{dH}{dt}-36H^4=0
\ee
then, for any given Hubble parameter $H(t)$, the particular solution 
\be\label{eq28} 
\psi(t)=6H(t)^2
\ee
satisfies this equation automatically. The corresponding scalar field is found as
\be\label{eq29}
\phi(t)=\sqrt{6}\int{H(t) dt}
\ee
The GB coupling is found by integrating Eq. (\ref{eq14c}) and the kinetic coupling from (\ref{eq14a}) becomes
\be\label{eq30}
F_1(t)=-\frac{1}{3H(t)^2}
\ee
therefore, for this particular choice of the GB coupling given by (\ref{eq14c}) with $g=-3/4$, we can give a dynamical interpretation to any given in advance Hubble parameter, in the frame of the present scalar field model with kinetic and GB couplings. In fact a Little Rip solution is characterized by a non singular energy density $\rho$ which increases with the scale factor $a$. Thus, for example taking the scale factor of the form \cite{paul}
\be\label{eq31}
a(t)=e^{\alpha(t)}
\ee
where $\alpha(t)$ is a non singular function of time, the Hubble parameter takes the form $H(t)=\dot{\alpha(t)}$. Then, the energy density becomes $\rho=3\dot{\alpha(t)}^2$, and the condition for increasing density is $\dot{H}=\ddot{\alpha}>0$, which automatically gives super-accelerating behavior. From Eq. (\ref{eq27}-\ref{eq29}) follows the expression for scalar field 
\be\label{eq32}
\phi(t)=\sqrt{6}\alpha(t)
\ee
The GB coupling is found as 
\be\label{eq33}
F_2(t)=-\frac{3}{4}\int{\frac{dt}{\dot{\alpha(t)}}}
\ee
and the kinetic coupling from (\ref{eq14a}) becomes
\be\label{eq34}
F_1(t)=-\frac{1}{3\dot{\alpha(t)}^2}
\ee
Of course, from this class of LR solutions we should pick up those that satisfy the restrictions imposed by astrophysical observational data. Particularly at current epoch, such models should give a behavior close to the $\Lambda$CDM.\\
The reconstruction considered here is based on the particular solution of the Eq. (\ref{eq27}) given by the Eq. (\ref{eq28}). Thus, for a suitable given $\alpha(t)$ we find the scalar field and the couplings $F_1$ and $F_2$ through Eqs. (\ref{eq32}-\ref{eq34}). Note that this reconstruction is applied to late time cosmological evolution. Let us consider the following late time dependence of the scale factor
\be\label{eq35}
a(t)=e^{\lambda t^n}
\ee
where $\lambda$ is a positive parameter and $n>1$. Then, $\alpha(t)=\lambda t^n$, $H=\dot{\alpha}=n\lambda t^{n-1}$ and $\phi=\sqrt{6}\lambda t^n$. From (\ref{eq33},\ref{eq34}) it follows
\be\label{eq36}
F_2=\frac{3}{4\lambda n(n-2)}t^{2-n},\,\,\,\,\ F_1=-\frac{1}{3\lambda^2 n^2}t^{2-2n}
\ee
expressed in terms of the scalar field, these couplings take the form 
\be\label{eq37}
F_2=\frac{3}{4\lambda n(n-2)}\left(\frac{\phi}{\sqrt{6}\lambda}\right)^{(2-n)/n},\,\,\,\,\, F_1=-\frac{1}{3\lambda^2 n^2}\left(\frac{\phi}{\sqrt{6}\lambda}\right)^{2(1-n)/n}
\ee
this completes the reconstruction. From (\ref{eq35}) and the expression for the Hubble parameter it follows that the effective density behaves as $\rho\propto (\ln a)^{2(1-1/n)}$, which for $n>1$ avoids the BR singularity and satisfies the condition for LR \cite{paul}, \cite{brevik} (i.e. $\ddot{\alpha}=\dot{H}>0$, producing super-acceleration, but the universe needs an infinite time to reach $\rho\rightarrow\infty$). Note that $n=1$ gives the de Sitter universe ($\rho=const.$), and when $n>>1$ the couplings behave as $F_1\sim \phi^{-2}$ and $F_2\sim \phi^{-1}$, like the couplings given in (\ref{eq26c}).

\section{Discussion}
In the frame of the scalar field model with kinetic and Gauss Bonnet couplings, we considered solutions to the dark energy problem that describe a quintom behavior, with and without BG singularities. According to this solution, it is possible to enter into the phantom phase without facing future BR singularities as the case described in Fig. 1; but there is also the possibility of getting into a BR singularity, depending on the values taken by the parameters, as shown in Fig. 2.  The new solution (\ref{eq15}) for the Hubble parameter evolves through the three phases: decelerated expansion, accelerated expansion and the final phantom phase with super accelerated expansion. The time of transitions can be set by adjusting the parameters of the solution. Particularly in the examples we considered here, at the current epoch the Universe is undergoing the transition to the phantom phase. The results presented in Figs. 1 and 2 indicate that knowledge of the asymptotic behavior of $w(t)$ (at $t \rightarrow\infty$) is insufficient to distinguish
models with a rip from models which are asymptotically de Sitter.
We have also studied another type of solutions called Little Rip, that cause the dissociation of structures in the Universe, due to the increasing behavior of the dark energy density. This is the same effect caused by Big Rip solutions, but without finite time singularities.\\
A remarkable aspect of the present model is that under the choice of the GB coupling $dF_2/dt=g/H$ with $g=-3/4$, the equation of motion is automatically satisfied for the particular form of the kinetic term as $\psi=6H^2$. For this particular solution, the Friedmann equation gives the kinetic coupling of the form $F_1=-1/(3H^2)$. This allows to reconstruct a class of cosmological scenarios explaining dark energy in the frame of the present model, as the examples considered above.

\end{document}